\documentclass[amsmath,amssymb,aps,pre,reprint, showpacs, superscriptaddress]{revtex4-1}
\usepackage{graphicx}
\usepackage{amsthm}
\usepackage{amsmath}
\usepackage{amssymb}
\usepackage{dsfont}
\usepackage{bm}
\usepackage{tikz}
\usepackage{tikz-3dplot}
\usepackage{epstopdf}
\usepackage{hyperref}
\usepackage{color}

\DeclareMathOperator{\Tr}{Tr}

\begin{document}
\title[]{Work and power fluctuations in a critical heat engine}
\author{Viktor Holubec}
\email{viktor.holubec@mff.cuni.cz}
\affiliation{ 
Institut f{\"u}r Theoretische Physik, 
Universit{\"a}t Leipzig, 
Postfach 100 920, D-04009 Leipzig, Germany
}
\affiliation{ 
 Charles University,  
 Faculty of Mathematics and Physics, 
 Department of Macromolecular Physics, 
 V Hole{\v s}ovi{\v c}k{\' a}ch 2, 
 CZ-180~00~Praha, Czech Republic 
}
\author{Artem Ryabov}
\affiliation{ 
 Charles University,  
 Faculty of Mathematics and Physics, 
 Department of Macromolecular Physics, 
 V Hole{\v s}ovi{\v c}k{\' a}ch 2, 
 CZ-180~00~Praha, Czech Republic 
}
\date{\today} 
\begin{abstract} 
We investigate fluctuations of output work for a class of Stirling heat engines with working fluid composed of interacting units and compare these fluctuations to an average work output. In particular, we focus on engine performance close to a critical point where Carnot's efficiency may be attained at a finite power as reported in [M. Campisi and R. Fazio, Nat. Commun. {\bf 7}, 11895 (2016)]. We show that the variance of work output per cycle scales with the same critical exponent as the heat capacity of the working fluid. As a consequence, the relative work fluctuation diverges unless the output work obeys a rather strict scaling condition, which would be very hard to fulfill in practice. Even under this condition, the fluctuations of work and power do not vanish in the infinite system size limit. Large fluctuations of output work thus constitute inseparable and dominant element in performance of the macroscopic heat engines close to a critical point.
\end{abstract}

\pacs{05.20.-y, 05.70.Ln, 07.20.Pe} 

\maketitle

\section{Introduction}

Heat engines and their performance trouble minds of 
engineers and physicists for more than two centuries \cite{Muller2007}.
One question is particularly exciting: Can efficiency of a
practical heat engine working between temperatures $T_h$
and $T_c$, $T_h>T_c$, reach the upper bound $\eta_C = 1-T_c/T_h$ found by 
Sadi Carnot \cite{Benenti2011,Allahverdyan2013,Proesmans2015,Brandner2015,Polettini2015,Shiraishi2017, Shiraishi2016}? 
Recent surprising theoretical breakthroughs suggest that 
this may be possible. The long lasting conviction that
Carnot's bound can be reached in a quasi-static (infinitely slow) 
limit only, and thus also at a vanishing power, has been proven unfounded. Two
ways how to reach $\eta_C$ at a finite power has been suggested.
First, the bound may be reached in the limit of infinitely fast
dynamics of the system \cite{Polettini2016,Lee2016,Holubec2017}. Second, the bound may be saturated in
heat engines based on systems working near critical point in case the 
heat capacity of the working medium scales in a suitable manner  
with the system size measured by the number of interacting particles, $N$ 
\cite{Campisi2016,Johnson2017}.

In this work, we will take a closer look on the second series of results.
Near the second order phase transition, space correlation functions of the system diverge and thus the system experiences large fluctuations. These fluctuations
allow the system working close to the critical point to reach
Carnot's bound at a finite power in the large $N$ limit. It is 
natural to ask whether also work fluctuations show critical behavior and, if so, whether these fluctuations do not dominate the average work produced by the engine.

In what follows, we derive concise formulas for work fluctuation, $\sigma^2_w$, 
and for relative work fluctuation (relative standard deviation), $f_w$, for the class of 
Stirling heat engines introduced in the inspiring study \cite{Campisi2016} and discuss their physical consequences. The results show that in the
leading order in the distance from Carnot's efficiency, $\Delta\eta = \eta_C - \eta$, 
the work variance is proportional to the heat capacity
\begin{equation}
\sigma^2_w = \left<w^2\right> - \left<w\right>^2 \propto C.
\label{eq:sigma1}
\end{equation}
Combining this result with the formula $\left<w\right>/\Delta\eta \propto C$ found by Campisi and Fazio \cite{Campisi2016} we obtain the following scaling of the relative work fluctuation
\begin{equation}
f_w = \frac{\sigma_w}{\left<w\right>} \propto \frac{1}{\sqrt{C}\Delta\eta}.
\label{eq:SNR1}
\end{equation}
Above, $C$ denotes the heat capacity of the working substance, $\left<w\right>$ the average work performed by the engine and $\left<w^2\right>$ the corresponding second moment.

Using these formulas, we discuss performance of heat engines working during the whole cycle in the vicinity of a critical point. Then the thermodynamic behavior of the engine is determined by a set of critical exponents of the engine working fluid \cite{Kiran2012, Campisi2016}. We find that such heat engines exhibit behavior which contradicts what is observed for standard non-interacting heat engines away from the phase transition. For the latter, the relative work fluctuation decreases as $f_w \propto 1/\sqrt{N}$ ensuring negligible fluctuations in the macroscopic limit. We find that the output work fluctuations for the critical heat engine increase with the number of interacting subsystems $N$ and diverge in the large system limit unless the strict scaling condition 
\begin{equation}
\left<w\right> \propto \frac{1}{\Delta \eta} \propto \sqrt{C} \propto N^{1/(2-\alpha)},\quad -\infty < \alpha < 2,
\label{eq:scaling_intro}
\end{equation}
is fulfilled. In this special case $f_w \propto 1$ and thus one obtains in the large system limit finite, but nonzero work fluctuations which are normally observed for microscopic heat engines only \cite{Crepieux2016,Holubec2014,Chvosta2010,Holubec2014a,Blickle2012,
Rana2014,Verley2014,Martinez2016,Sinitsyn2011,Rossnagel2016,Basu2016}. 

In order to achieve the favorable scaling (\ref{eq:scaling_intro}) in an actual heat engine, one must know the critical exponent $\alpha$ exactly. Such a precision is impossible to achieve in practical experiments or numerical simulations. Our findings thus practically close the way to construct a standard macroscopic critical heat engine operating at Carnot's efficiency at a finite average power with finite fluctuation. 

This, however, does not mean that the idea of a critical heat engine is completely useless. One can for example construct a non-standard heat engine with a feedback mechanism which would harvest the positive work fluctuations from the system. On the other hand, the sensitivity of the critical heat engine to the scaling of the mean output work can be used as a basis of a novel experimental technique for precise measurement of the critical exponent $\alpha$. Last but not least, if the scaling would be close enough to Eq.~(\ref{eq:scaling_intro}), one can construct a heat engine working at a large efficiency (although smaller than $\eta_C$) and a large output power with acceptable fluctuation by choosing a suitable finite value of $N$.

\section{Model}

\begin{figure}
\centering
\includegraphics{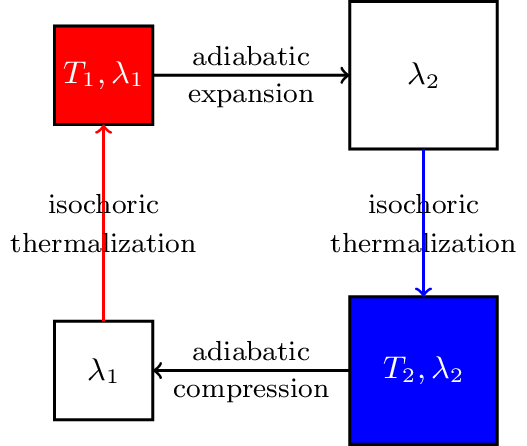}
\caption{(Color online) The Stirling cycle. The engine exchanges heat with
reservoirs (performs work on the environment) during the isochoric (adiabatic) 
branches only.}
\label{fig:cycle}
\end{figure}

Consider the Stirling cycle performed by a medium described by the
Hamiltonian 
\begin{equation}
H(t) = \lambda(t)K,
\label{eq:Hamiltonian}
\end{equation}
where $K$ is an $N$-particle Hermitian operator and $\lambda(t)$ is a parameter which can be 
varied in time by the experimentalist. We assume that the Hamiltonian is
confining so for any reasonable $\lambda=\lambda(t)>0$ there exists Gibbs equilibrium 
state $P_{eq}(\beta \lambda) = \exp(-\beta \lambda K)/\Tr[\exp(-\beta \lambda K)]$,
$\beta = (k_B T)^{-1}$. By varying the parameter $\lambda(t)$ we change the ``volume'' of the whole $N$-particle system. We are interested in the collective response of the whole system to the driving which may lead to an enhanced performance of the heat engine as compared to the system composed of $N$ non-interacting particles. This model was first introduced in Ref.~\cite{Campisi2016}.

The considered Stirling cycle is depicted in Fig.~\ref{fig:cycle}. It
is composed of two adiabatic processes during which the system distribution 
does not change and the system performs work on its surroundings and two very 
slow isochoric processes where the system eventually equilibrates. During the 
first adiabatic process, the Hamiltonian changes from $\lambda_1 K$ to $\lambda_2 K$, 
and the system is initially in equilibrium with the bath at the temperature $T_1$. 
Along the second adiabatic process, the Hamiltonian 
changes back from $\lambda_2 K$ to $\lambda_1 K$ and the system starts from equilibrium at
the temperature $T_2$. We assume that $\lambda_2<\lambda_1$, and that $T_1>T_2$. Then the 
adiabatic process at $T_1$ mimics adiabatic expansion (the confinement $\lambda$ changes 
form stronger to weaker) and vice versa for the adiabatic process at $T_2$. Assuming further
$\lambda_2/\lambda_1 > T_2/T_1$ the system operating in a time-periodic steady state
performs on average more work during the expansion than it consumes
during the compression and thus it operates as a heat engine \cite{Campisi2016}.

Let us now investigate the output work of the engine. In order to do that, it is favorable
to work in a specific basis $\left|x\right>$, where $x$ stands for  
a set of independent internal degrees of freedom of the system. If we denote
as $K(x) = \left<x\right|K\left|x\right>$ the corresponding elements of the 
Hamiltonian (\ref{eq:Hamiltonian}), the stochastic work done by the system during 
one cycle reads
\begin{equation}
w(x_1,x_2) = (\lambda_1 - \lambda_2)(K(x_1) - K(x_2)),
\label{eq:random_work}
\end{equation}
where $x_1$ ($x_2$) is the state of the system during the first (second) adiabat. Because the 
isochoric processes allow the system to relax to equilibrium, the variables $x_1$ and $x_2$ 
are independent and the probability density of the work $w$ is given by the formula
\begin{multline}
\rho(w) = \int\int dx_1 dx_2 \,\delta[w - w(x_1,x_2)] \\
\times \frac{\exp[-\beta_1\lambda_1 K(x_1)]}{Z_1}
\frac{\exp[-\beta_2\lambda_2 K(x_2)]}{Z_2},
\label{eq:prob_work}
\end{multline}
where $\delta(\bullet)$ denotes the Dirac's $\delta$-function. The function (\ref{eq:prob_work}) 
allows us to calculate all moments of the output work.

The mean work along one cycle reads
\begin{equation}
\left<w\right> = \int dw\, w \rho(w) = (\lambda_2 - \lambda_1) \left(\left<K\right>_{\beta_2 \lambda_2} - \left<K\right>_{\beta_1 \lambda_1}\right)
\label{eq:mean_work}
\end{equation}
and for the work fluctuation we get
\begin{multline}
\sigma_w^2 = (\lambda_2 - \lambda_1)^2 \left[\left( 
\left<K^2\right>_{\beta_2 \lambda_2} - \left<K\right>_{\beta_2 \lambda_2}^2
\right) \right. \\ \left. 
+ \left( 
\left<K^2\right>_{\beta_1 \lambda_1} - \left<K\right>_{\beta_1 \lambda_1}^2
\right)
\right].
\label{eq:mean_squared_work}
\end{multline}
Above, $\left<K\right>_{\beta_1 \lambda_1}$ denotes average with respect to the distribution
$\exp[-\beta_1\lambda_1 K(x)]/Z_1$ and similarly for $\left<K\right>_{\beta_2 \lambda_2}$. The work fluctuation determine the number of cycles for which the output work of the heat engine must be measured in order to get a reliable prediction of the mean output work. According to the central limit theorem, the fluctuation of the predicted average work after a large enough number $N_m$ of measurements behaves as $\sigma_w^2/N_m$. Large fluctuation $\sigma_w^2$ implies that one have to measure for a long time to get a reliable prediction of the mean work. 

\section{Near Carnot's bound}

The heat accepted by the system from the hot bath is given by
$\left<q_{in}\right> = \lambda_1(\left<K\right>_{\beta_1 \lambda_1} - \left<K\right>_{\beta_2 \lambda_2})$
and thus the efficiency of the Stirling engine reads \cite{Campisi2016}
\begin{equation}
\eta = \frac{\left<w\right>}{\left<q_{in}\right>} = 1 - \frac{\lambda_2}{\lambda_1} < 1- \frac{\beta_1}{\beta_2} = \eta_C.
\label{eq:efficiency}
\end{equation}
As outlined in the Introduction, the main aim of the present paper is to investigate fluctuations of the output work of the critical heat engine operating close to Carnot's bound. The distance from the Carnot's bound is for the present model given by \cite{Campisi2016} 
\begin{equation}
\Delta \eta = \eta_C-\eta = \frac{\lambda_2}{\lambda_1} - \frac{\beta_1}{\beta_2} > 0.
\label{eq:distance_from_etaC}
\end{equation}
In what follows, we will study the behavior of work fluctuation $\sigma_w^2$ and relative work fluctuation $f_w$ up to the first order in $\Delta \eta$.

In Ref.~\cite{Campisi2016} it was shown that for small
$\Delta \eta$ the ratio of mean work and the distance from Carnot's bound is proportional to the
heat capacity of the system:
\begin{equation}
\Pi = \frac{\left<w\right>}{\Delta\eta} \propto C_{\beta_1 \lambda_1},
\label{eq:ratio_w}
\end{equation}
where $C_{\beta_1 \lambda_1} = -k_B\beta_1^2 dU_{\beta_1\lambda_1}/d\beta_1$ denotes the heat
capacity and $U_{\beta_1\lambda_1} = \lambda_1 \left<K\right>_{\beta_1 \lambda_1}$ the 
internal energy of the system.

From the formula (\ref{eq:mean_squared_work}) for the work fluctuation it follows that $\sigma_w^2$ is determined by
the equilibrium fluctuations of the internal energy of the system 
$\sigma^2_{u_{\beta_i\lambda_i}} =\lambda_1^2\left<K^2\right>_{\beta_i\lambda_i} - U^2_{\beta_i\lambda_i}$, $i = 1,2$, 
which are in turn proportional to the heat capacities $\sigma^2_{u_{\beta_i\lambda_i}} = \beta_1^{-2} C_{\beta_i \lambda_i} /k_B$, $i = 1,2$. Altogether we get 
\begin{equation}
\sigma_w^2 = \frac{(\lambda_2 - \lambda_1)^2}{k_B}
\left(
\frac{1}{\beta_2^2\lambda_2^2} C_{\beta_2 \lambda_2} + \frac{1}{\beta_1^2\lambda_1^2} C_{\beta_1 \lambda_1}
\right).
\label{eq:work_fluctuations}
\end{equation}
From Eq.~(\ref{eq:distance_from_etaC}) it follows that $\beta_2\lambda_2 = \beta_2\lambda_1\Delta\eta + \beta_1\lambda_1$, and thus, up to the leading order in $\Delta \eta$, the work fluctuation can be further rewritten as
\begin{equation}
\sigma_w^2 \approx 2\frac{(\lambda_2 - \lambda_1)^2}{k_B \beta_1^2 \lambda_1^2}
C_{\beta_1 \lambda_1} = 2\frac{\eta^2}{k_B \beta_1^2}
C_{\beta_1 \lambda_1} \propto C_{\beta_1 \lambda_1}.
\label{eq:work_fluctuatin_final}
\end{equation}
The work variance hence scales in the same way as $\Pi$. The relative work fluctuation then 
scales as
\begin{equation}
f_w \propto \frac{1}{\sqrt{C_{\beta_1 \lambda_1}} \Delta \eta}.
\label{eq:relative_work_fluctuation}
\end{equation}
Introducing duration of one period of the cycle $\tau$ and defining the fluctuating output power as $p(x_1,x_2) = w(x_1,x_2)/\tau$ we find that the formula (\ref{eq:relative_work_fluctuation}) describes also the relative fluctuation of the output power: $f_p = \sqrt{\left<p^2\right> - \left<p\right>^2}/\left<p\right> = f_w$.
Equations (\ref{eq:work_fluctuatin_final}) and (\ref{eq:relative_work_fluctuation}) constitute our
first main result. The next sections emphasize their significance for the engine performance.

\section{Near the phase transition}

The engine working fluid consists of $N$ interacting subunits. In Eq.~(\ref{eq:Hamiltonian}), the interaction is incorporated in the $N$-particle operator $K$.
Without the interaction, the thermodynamic quantities in question would follow the well known scaling
behavior: $C_{\beta_1\lambda_1} \propto \left<w\right> \propto \sigma^2_w \propto N$ and $\Delta\eta \propto 1$.
Thus, by enlarging the number of subunits, one would not get any interesting gain in the engine
performance. The main idea of Ref.~\cite{Campisi2016} is that the above simple scaling can change if the subunits interact and, for example, the engine works in the neighborhood of a critical 
point (phase transition). Let us now use the formulas (\ref{eq:work_fluctuatin_final}) and (\ref{eq:relative_work_fluctuation}) to uncover another side of this idea omitted in Ref.~\cite{Campisi2016}. The following analysis shows that, together with the enhanced scaling of the mean output work and efficiency found in Ref.~\cite{Campisi2016}, the engine working close to a critical point inevitably exhibits also unfavorable scaling of work fluctuations.

Near the phase transition, the system behavior can be described in a very general fashion using the
finite size scaling theory \cite{Fisher1972, Suzuki1977,Campisi2016}. Let us denote as $T_c$ the temperature 
of the phase transition in question, $\theta = |T-T_c|/T_c$ the distance from the critical temperature and
$d$ the dimension of the system. When reaching the critical temperature ($\theta\to 0$) in the thermodynamic limit, 
the heat capacity $C$ diverges with the exponent $\alpha > 0$ as $C \propto \theta^{-\alpha}$ 
and the correlation length $\xi$  diverges with the exponent $\nu > 0$ as $\xi \propto \theta^{-\nu}$. 
For finite size systems operating close to the critical point, the heat capacity develops a peak with height $C$ 
and width $\delta$ which scale as $\log C / \log N \propto  1 + \alpha/(d\nu) = 2/(2-\alpha)$ and $\log \delta / \log N = - 1/(d\nu) = - 1/(2-\alpha)$, respectively.
Here we have used the formula $\nu d = 2-\alpha$ \cite{Huang2009}.

As explained in Ref.~\cite{Campisi2016}, the scaling of the heat capacity has important consequences for the 
validity of approximations used in the derivation of Eqs.~(\ref{eq:ratio_w}), (\ref{eq:work_fluctuatin_final}) and (\ref{eq:relative_work_fluctuation}). In deriving the formula (\ref{eq:ratio_w}) one assumes that
the internal energy $U_{\beta_2\lambda_2}$ can be approximated by the first order Taylor expansion around $\beta_1 \lambda_1$. Similarly, in the derivation of Eq.~(\ref{eq:work_fluctuatin_final}) we assume that the heat capacity $C_{\beta_2\lambda_2}$ can be approximated by zeroth order expansion around $\beta_1 \lambda_1$. More
precisely, one assumes that $\Pi=\left<w\right>/\Delta\eta \approx \partial \left<w\right>/\partial \Delta\eta|_{\Delta\eta = 0}$, $U_{\beta_2\lambda_2} = U_{\beta_2\lambda_1\Delta\eta + \beta_1\lambda_1} \approx U_{\beta_1\lambda_1} + dU_x/dx|_{x=\beta_1\lambda_1} \beta_2\lambda_1\Delta\eta$ and $C_{\beta_2\lambda_2} = C_{\beta_2\lambda_1\Delta\eta + \beta_1\lambda_1} \approx C_{\beta_1\lambda_1}$.
Assuming that the inverse critical temperature of the Hamiltonian $K$ is given by $\beta_1\lambda_1$, the expression $\beta_2\lambda_1\Delta\eta = \beta_1\lambda_1 - \beta_2\lambda_2$ measures the distance from the critical point. The region where the internal energy can be approximated by the linear expression above for a given $N$ is given by the width $\delta$ of the heat capacity peak. This means that $\beta_2\lambda_1\Delta\eta$ must shrink with increasing $N$
in the same manner as $\delta$ or faster in order to secure the validity of Eq.~(\ref{eq:ratio_w}). We arrive at the condition 
\begin{equation}
\Delta \eta \propto N^{-\gamma},\quad \gamma \ge 1/(2-\alpha) \ge 0
\label{eq:peak_condition}
\end{equation}
which also secures validity of the above approximation of the heat capacity.

Physically, the validity of the linear approximation means that the engine operates in the neighborhood of the critical point during the whole operation cycle, i.e. $\beta_2\lambda_2$ is ``close'' to $\beta_1\lambda_1 = \beta_c$. The size of this neighborhood shrinks with increasing number of subunits $N$ proportionally to $\delta$ and hence the distance from the phase transition, measured by $\Delta\eta$, must shrink in the same manner or faster in order to keep the engine close to the phase transition for any $N$.

Inserting the scaling (\ref{eq:peak_condition}) of $\Delta \eta$ together
with the scaling of heat capacity into the formula (\ref{eq:relative_work_fluctuation}) for the relative work fluctuation $f_w$, we find
\begin{equation}
f_w \propto N^{-1/(2-\alpha)+\gamma}.
\label{eq:fw_scaling}
\end{equation}
In order to have finite $f_w$ in the large $N$ limit we thus arrive at the condition
$1/(2-\alpha) \ge \gamma$. Altogether, if one asks both for the validity of the linear
approximation described above and for a nonzero relative work fluctuation, the critical exponent must obey the strict condition $1/(2-\alpha) = \gamma$
which leads to the scaling 
\begin{equation}
\Delta \eta \propto \frac{1}{\sqrt{C}} \propto N^{-1/(2-\alpha)}
\label{eq:delta_eta_OK_scaling}
\end{equation}
of the distance from Carnot's efficiency. This observation constitutes our second main result. Let us stress that for non-interacting subsystems
the relative work fluctuation scales as $f_w \propto 1/\sqrt{N}$, while for the present critical heat engine we obtain at best $f_w \propto 1$ if we want to work in the linear regime of the internal energy. Machines working close to a critical point can thus be expected to exhibit large work fluctuations even in the large system limit. 

Assuming the above strict scaling (\ref{eq:delta_eta_OK_scaling}), we obtain from Eq.~(\ref{eq:ratio_w}) that the mean output work must behave as
\begin{equation}
\left<w\right> \propto \sqrt{C} \propto N^{1/(2-\alpha)}.
\label{eq:W_OK_scaling}
\end{equation}
Such scaling of output work can be achieved for example by fixing $\lambda_1$ and carefully choosing the parameter $\lambda_2$ for each value of $N$ such that the formula $\left<w\right> \propto \sqrt{C}$ is fulfilled \cite{Campisi2016}. In order to do this, one should however know the exact value of the critical exponent $\alpha$. This represents a thin bottle neck which may be really hard to surpass as discussed at the end of the present section. Nevertheless, let us now investigate the performance of hypothetical heat engines where the scaling (\ref{eq:W_OK_scaling}) can be achieved in practice.

According to the dynamical finite-size scaling theory, close to a critical point the relaxation time to equilibrium scales with the number of subsystems as $\tau_R \propto N^{z/d}$, where $z$ can be both positive (critical slowing down) and negative (critical speeding up) \cite{Campisi2016}. It is thus reasonable to assume that the time needed to perform the isochoric branches of the Stirling cycle, where the systems thermalizes with the individual heat reservoirs, scales as $N^{z/d}$. If we further assume that the adiabatic branches are performed infinitely fast,
the total duration of the cycle $\tau$ would scale as $\tau \propto N^{z/d}$. This is the shortest possible duration of the cycle which maximizes the power output. Assuming that the duration of the adiabatic branches scales with $N$, an arbitrary scaling of the cycle duration of the form $\tau \propto N^{\zeta/d} \ge N^{z/d}$ can be obtained.

Under the assumption of the fastest possible cycle ($\tau \propto N^{z/d}$) the average output power of the engine scales as
\begin{equation}
\left<p\right> = \frac{\left<w\right>}{\tau} \propto \frac{\sqrt{C}}{\tau} \propto N^{1/(2-\alpha) - z/d}.
\label{eq:P_OK_scaling}
\end{equation}

The formula $d\nu = 2-\alpha$ implies that $-\infty < \alpha < 2$. In the lucky case of Eq.~(\ref{eq:delta_eta_OK_scaling}), the work and power thus scale at worst ($\alpha \to -\infty$) as $\left<w \right> \propto 1$ and $\left<p \right> \propto N^{-z/d}$, respectively. The scaling of work is better than in the situation when the $N$ subsystems do not interact ($\left<w\right> \propto N$) whenever $\alpha>1$. A gain in power is obtained only if the further condition $1/(2-\alpha) - z/d > 1$ leading to
\begin{equation}
\frac{z}{d} < \frac{\alpha-1}{2-\alpha}
\label{eq:z_OK}
\end{equation}
is fulfilled. For example, for the 3D Ising model we have $\alpha \approx 0.12$ \cite{Pelissetto2002} and $z \approx 2.35$ \cite{Matz1994} leading to $\left<w\right> \propto N^{0.53}$ and $\left<p\right> \propto N^{-0.25}$. 
Another example is the exotic substance $\rm Dy_2 Ti_2 O_7$
which posses the rather large coefficient $\alpha \approx 0.38$ \cite{Higashinaka2004} and exhibits the critical speeding up ($\nu z \approx -0.7$ \cite{Grams2014} and thus $z \approx -1.3$). In three dimensions, these exponents lead to the scaling $\left<w\right> \propto N^{0.62}$ and $\left<p\right> \propto N^{1.05}$. In this case, the critical speeding up further allows to reduce the work fluctuation by averaging the work output over a large number of cycles in a short time. Indeed, the amount of cycles per unit time scales as $N_m = 1/\tau \propto N^{-z/d} = N^{0.43}$ and thus the relative variance $f_w/\sqrt{N_m}$ of the output work averaged over the unit time vanishes with $N$ as $f_w/\sqrt{N_m} \propto N^{z/(2d)} = N^{-0.22}$. Let us note that in a more realistic situation where the total cycle time scales as $\tau \propto 1$ and such averaging is not possible, the heat engine based on $\rm Dy_2 Ti_2 O_7$ still reaches useful output power which scales as $\left<p\right> \propto N^{0.62}$ (scaling of other variables do not change).


Our analysis of work fluctuations thus leaves open the way to construct a classical macroscopic heat engine working at Carnot's efficiency at a nonzero average power suggested by Campisi and Fazio as long as the scaling (\ref{eq:delta_eta_OK_scaling}) of the output work can be experimentally achieved. However, to do this, one should know the exact value of the critical exponent $\alpha$. Obtaining an exact value of a physical parameter is impossible to achieve both in experiments and in numerical simulations and thus it is our opinion that it would be nearly impossible to achieve the scaling (\ref{eq:delta_eta_OK_scaling}) in practice. Even very small discrepancy in the used $\alpha$ would lead to a different scaling. Whenever the work would scale slower than $\sqrt{C}$ the heat engine would not work in the critical regime as discussed above and thus the analysis using the critical exponents would not be correct. On the other hand, whenever the work would scale faster than $\sqrt{C}$ the output work and power would be dominated by fluctuations rendering the average work/power output of the machine useless (unmeasurable) for large $N$, where Carnot's efficiency is achieved. One may argue that the scaling $\tau \propto N^{z/d}$ of the total cycle time may allow us to average out the large work fluctuations in finite time as mentioned above for $\rm Dy_2 Ti_2 O_7$. However, also this scaling is beyond the reach of experiments where the total cycle time is always bounded from bellow and thus it scales rather as $\tau \propto 1$.

Instead of utilizing the average work/power of such a machine in the large $N$ limit, one should rather focus on harvesting the positive power fluctuations as can be achieved in feedback driven systems. The strong sensitivity of the critical heat engine to the scaling of the output work can be also used as a basis of a novel experimental technique for precise measurements of the critical exponent $\alpha$.

To conclude, the critical heat engines can be in practice used as standard heat engines only if one does not pursue the large $N$ limit too far. In such a case, choosing the scaling of the output work close enough to (\ref{eq:W_OK_scaling}) will result in a weak divergence of the relative power fluctuation with $N$ (for example taking $\gamma = 1/(2-\alpha) + \epsilon$ will lead to the scaling $f_w \propto N^\epsilon$, the parameter $\epsilon>0$ can be related to the uncertainty in determination of the exponent $\alpha$). One can then construct a heat engine operating close to Carnot's efficiency at a large average power output by choosing $N$ small enough that the power fluctuations are acceptable for the respective application (for example the measurement time available for averaging out the fluctuations).

\section{Conclusion and outlook}

We have studied the class of Stirling heat engines described in Fig.~\ref{fig:cycle}.
In case the heat engine is based on $N$ interacting subsystems and operates during the whole cycle close to a critical point, we have found
the strict condition (\ref{eq:scaling_intro}) under which the engine can at the 
same time reach Carnot's efficiency and a finite relative work fluctuation in the large $N$ limit. In the unlikely case when this condition can be fulfilled, the relative work fluctuation does not scale with $N$ and thus it remains finite in the large $N$ limit. In all other cases, the work fluctuations dominate the average output work of the engine and renders the actual output work/power of such engines nearly unpredictable. Thus the device can not work as a classical heat engine. One should rather supplement the it with a feedback mechanism which would harvest positive work fluctuations. The strong sensitivity of work fluctuations to the scaling of the output work in heat engines working close to a critical point may be used to a novel experimental technique for precise measurements of the critical exponent $\alpha$.

The restriction (\ref{eq:scaling_intro}) results from the necessity to work within the linear regime around the critical point in the derivation of the results. The situation may change when different Hamiltonians and/or thermodynamic
cycles are considered. Furthermore, different results can be obtained outside the linear regime, i.e. for heat engines which do not operate close to the critical point during the whole cycle. 
One can also think about different ways of achieving suitable scaling of thermodynamic 
variables in question, namely of the mean work, distance from Carnot's bound and relative work/power fluctuation (see for example the Ref.~\cite{Polettini2016} where such a scaling is discussed for a quantum dot and Ref.~\cite{Holubec2017} where a scaling leading to Carnot's efficiency at a nonzero power and finite power fluctuation is presented for a Brownian heat engine). Finally, as already highlighted in Ref.~\cite{Campisi2016}, it is not necessary to 
consider the limit of large $N$ in order to get better performance with respect to non-interacting systems. 

Although Carnot's efficiency is achieved in the large $N$ limit only, also the power fluctuations diverge only in this limit. Hence, by choosing a finite $N$, it is possible to achieve a suitable trade-off between large efficiency, average output power and acceptably large power fluctuation. For a small enough system, one can further in detail optimize interactions between the individual subunits in order to get the desired engine behavior. Such detailed optimization would be indeed impossible for systems composed of many subunits.

The take-home message of the present work is very simple:
when studying performance of heat engines in exotic regimes of operation where thermodynamic quantities such as correlation functions diverge, one should always address both the mean values and the fluctuations. Only then one can be sure that the average values can be observed experimentally 
and they wont get lost in a deep forest of randomness.

\begin{acknowledgments}
The authors thanks M. Campisi and N. Shiraishi for inspiring discussions. Support of the work by the COST Action MP1209 and by the Czech Science Foundation (project No. 17-06716S) is gratefully acknowledged. VH in addition gratefully acknowledges the support by Alexander von Humboldt agency.
\end{acknowledgments}

%

\end{document}